\documentstyle[12pt,fleqn]{article}
\newcommand{\beq}{\begin{equation}}
\newcommand{\eeq}{\end{equation}}
\def\hg#1{{\hat g}_{#1}}
\def\p#1#2{{\partial #1\over\partial #2}}
\newcommand{\beqa}{\begin{eqnarray}}
\newcommand{\eeqa}{\end{eqnarray}}

\begin{document}

\centerline{\large\bf
Anisotropic Inflation from Extra Dimensions}

\vspace{.5truecm}
\centerline{Marco Litterio$^{1,\star}$,  Leszek M.~Soko{\l}owski$^2$,
Zdzis{\l}aw A.~Golda$^{2}$}
\centerline{Luca Amendola$^3$ and Andrzej Dyrek$^4$}
\vspace{.3truecm}
{\footnotesize
\centerline{$^1${\em Istituto Astronomico, Universit\`a di Roma ``La Sapienza",
		Via Lancisi 29, I-00161 Roma, Italy}}
\centerline{$^2${\em Astronomical Observatory, Jagellonian University,
		Orla 171, 30-244 Krak\'ow, Poland}}
\centerline{$^3${\em Osservatorio Astronomico di Roma, Viale del Parco Mellini
84, I-00136 Roma, Italy}}
\centerline{$^4${\em Institute of Physics, Jagellonian
University, Reymonta 4, 30-059 Krak\'ow, Poland}}
}
\vspace{1.truecm}
{\small PACS: 98.80 Cq, 04.50+h

}
\vspace{.5truecm}
\baselineskip 19pt
\begin{abstract}
Vacuum multidimensional cosmological models with internal spaces
being compact $n$-dimensional Lie group manifolds are
considered. Products of 3-spheres and $SU(3)$ manifold (a novelty
in cosmology) are studied. It turns out that the dynamical evolution of
the internal space drives an accelerated expansion of the
external world (power law inflation). This generic solution
(attractor in a phase space) is determined by the Lie group
space without any fine tuning or arbitrary inflaton potentials.
Matter in the four dimensions appears in the form of a number of
scalar fields representing anisotropic scale factors for the
internal space. Along the attractor solution the volume of the
internal space grows logarithmically in time. This simple and
natural model should be completed by mechanisms terminating the
inflationary evolution and transforming the geometric scalar
fields into ordinary particles.
\end{abstract}
\vfill
{\footnotesize
$\star$ Corresponding author; e-mail address: litterio@astrom.astro.it;
fax: 39-6-4403673
}
\newpage

The cosmological evolution of the universe must be driven by some
kind of matter. If the large scale geometry of our spacetime
is described by isotropic
Friedmann-Lema\^{\i}tre-Robertson-Walker (FLRW)  models, as they
best fit to the observational data, the galactic matter (and
other, yet undetected, forms of matter) cannot be viewed as test
particles but as a self-gravitating fluid. In fact, the vacuum
FLRW models are either inconsistent (spatially closed models)
with Einstein field equations without cosmological constant or
merely describe flat Minkowski space (open and flat models). In
the early universe its self-gravitating contents form a
structureless plasma of all elementary particles present in the
standard model (and possible exotic particles beyond the model)
and the evolution rate and physical processes occurring in a
given era depend on which particle species dominates.

It is widely believed that the standard physical cosmology
should be modified in the earliest eras by the concept of
inflationary evolution \cite{infl}. Inflationary (i.e.
accelerated) expansion is possible due to the existence of
large negative pressure in the gravitating source, therefore it
cannot be driven by a relativistic gas of the particles in
thermal equilibrium. Only scalar fields can provide negative
pressure. However the Higgs scalar of the standard model is
inappropriate to this aim. It is necessary to introduce {\em ad
hoc} a special {\em inflaton\/} field playing otherwise no role
in particle physics and this is why there have been proposed so
many mechanisms of inflation, none of which is well grounded in
fundamental physics.

Scalar fields appear in a natural way in higher-dimensional
physics. Yet it is rather little known what particles live in
higher dimensions. It is therefore interesting to assume that
the primordial universe was $d=4+n$-dimensional anisotropic and
empty. All its spatial dimensions evolved as being dynamical
degrees of freedom under influence of vacuum gravity. The total
spacetime was split into a product of four-dimensional
spacetime ${\cal M}$ named ``external space" (which has then evolved
into the world we see around us) and an $n$-dimensional closed
space ${\cal B}$ (``internal space") with a Riemanian metric. The
splitting allows for performing dynamical reduction of the full
theory to the low-energy physics occurring in the external space
${\cal M}$; the extra dimensions forming the internal space provide
then a number of scalar fields in ${\cal M}$ and this ``matter" of
geometrical origin acts as a source for $4$-dimensional gravity.
We find this model as natural, simple and elegant. It is
conceivable that the geometrical matter is later turned in some
process into ordinary particles of the standard model. We shall
not address here the latter problem.

In this letter we present results of qualitative analysis of
cosmological evolution of the external space which is driven by
$n$ scalar fields representing anisotropic evolution of $n$
internal dimensions. We are interested both in
isotropization process of these dimensions and in possible
inflation of the ordinary 3 external dimensions. We find as a
general rule that initial anisotropy of the internal space
diminishes in time and for some cases vanishes at all.
Inflationary solutions are also generic (attractor solutions)
and are usually associated with a mild, acceptable growth of the
internal space. Yet the graceful exit problem remains open.

The total spacetime $\cal P$ is taken as a ``warped" metric product
\cite{AA} ${\cal P}={\cal M}\times {\cal B}$, $\dim {\cal M}=4$ and
$\dim {\cal B}=n$. In general
the internal space $\cal B$ can evolve anisotropically since its
metric can depend on the external coordinates $x^\mu$,
$\mu=0,\ldots,3$. The components of the metric thus generate
$\frac{1}{2}n(n+1)$ scalar fields in ${\cal M}$. As $\cal B$ we choose a
compact Lie group manifold with a diagonal metric (giving rise
to $n$ scalars).

The metric tensor for ${\cal P}$ is then
\beq\label{eq1}
\hg{AB}(x,y)=\left(\begin{array}{cc}
{e}^{-nF(x)}g_{\alpha\beta}(x)    &0\\
0                               &L^2e^{2(F(x)+\phi_a(x))}\delta_{ab}
\end{array}\right),
\eeq
where $h_{ab}= L^2{e}^{2(F+\phi)}\delta_{ab}$,
$a,b=1,\ldots,n$, is a metric on $\cal B$ in a right-invariant
vector basis. $h_{ab}$ must be independent of internal
coordinates $y^a$. Here $\phi_a$ are the {\em relative
anisotropic deformation factors\/} subject to the linear
constraint $\sum_{a=1}^n\phi_a=0$; $L{e}^F$ is the mean
radius of the internal space and $F$ is the {\em dilaton\/}
field. Note that the {\em physical\/} metric field for ${\cal M}$ is
$g_{\alpha\beta}$ and not
$\hg{\alpha\beta}={e}^{-nF}g_{\alpha\beta}$, as required by
the fact that we have used the warped metric product (see for
details, \cite{DD}, cf. also \cite{Maeda,SC,Cho19871992}).

Since we assume Einstein gravity in the vacuum total spacetime,
the action is
\beq
S=\int_{\cal B}\!\!{\rm d}^ny\int\!\!{\rm d}^4x\,\sqrt{-\hat{\bf g}}\,
\Big[{\hat R}(\hat{\bf g})
-2\Lambda\Big],
\eeq
the presence of a (positive) cosmological constant is essential
for our analysis. Although the fields $F$ and $\phi_a$ have
direct geometrical meaning, after dimensional reduction they
give rise to non-standard kinetic terms in the effective action. In
order to give the action the standard form one makes linear
transformations to new fields, $P\equiv \alpha F -q$,
$\phi_a\equiv \sum_{i=1}^{n-1}A_{ai}\psi_i$ with $\alpha$, $q$ and $A_{ai}$
appropriately chosen constants. Then
\beq\label{eq3}
S=\int\!{\rm d}^4x\,\sqrt{-{\bf g}}\,\Big[R({\bf g}) -2\Lambda_4
-P_{,\mu}P^{,\mu} -\sum_{i=1}^{n-1}\psi_{i,\mu}\psi_{i}^{,\mu}
-2V(P,\psi)\Big],
\eeq
where $R({\bf g})$ is the curvature scalar for
$g_{\alpha\beta}(x)$, $\Lambda_4\equiv 2\Lambda/(n+2)>0$ and
\beq\label{eq4}
V(P,\psi)=\Lambda\Big[{e}^{-\beta P} -\mu{e}^{-\gamma P}R_{\cal B}
-\frac{2}{n+2}\Big]
\eeq
is an effective potential generated by the curvature scalar
$R_{\cal B}(\psi)$ of the internal space. Here
$\beta\equiv\sqrt{2n/(n+2)}=n/\alpha=2/\gamma$ and $\mu\equiv
\beta^2/2R_0$ with $R_0\equiv R(\phi_a=0)$. The physical system
in four dimensions consists of the metric $g_{\mu\nu}$ and the
``geometric matter": the dilaton $P$ and $n-1$ independent
deformation factors $\psi_i$, $i=1,\ldots,n-1$. The field
equations take on then the standard form
\beq\label{eq5}
G_{\mu\nu}+\Lambda_4g_{\mu\nu} = T_{\mu\nu}(P) +\sum_{i=1}^{n-1}
T_{\mu\nu}(\psi_i) - V(P,\psi)g_{\mu\nu},
\eeq
\beq\label{eq6}
\Box P=\p{V}{P}~~\mbox{and}~~\Box\psi_i=\p{V}{\psi_i}
\eeq
with
\beq\label{eq7}
T_{\mu\nu}(f)\equiv f_{,\mu}f_{,\nu}-\frac{1}{2}g_{\mu\nu}f_{,\alpha}
f^{,\alpha}.
\eeq
It is clear that the effective potential $V(P,\psi)$ fully
characterizes the dynamics of the system and in the following we
will discuss qualitative behaviour of the system employing the
potential instead of searching for exact solutions eqs.
(\ref{eq5})-(\ref{eq6}).

It turns out that Einstein field equations in $d$ dimensions are
compatible with the metric (\ref{eq1}) only if the Ricci tensor
for $h_{ab}$ on $\cal B$ is diagonal. This condition for compact
non-abelian Lie groups with diagonal metrics holds e.g. for the
$SU(3)$ space and products of 3-spaces (and is trivially
satisfied for the abelian flat group space, $n$-torus).

Consider first the $SU(3)$ group manifold. Although it has no
direct geometrical interpretation, its significance for particle
physics makes it an interesting model for the higher-dimensional
theory. As far as we know this space has never been investigated
in this context nor in cosmology. Requiring the Ricci tensor
for $SU(3)$ to be diagonal and taking into account the
constraint $\sum_a\phi_a=0$ one finds that there are at most
five independent deformation factors $\psi_i$ $(n=8)$. For the
external spacetime metric we take the simplest case --- flat
FLRW model. Even then generic dynamical system with the six
scalar fields is intractable. Fortunately, additional symmetries
of the system can be used to simplify the problem and it turns
out that the most symmetric case already contains all the
relevant features of the whole system and thus there is no loss
of generality in studying it.

The most symmetric $SU(3)$ case, compatible with the field
equations, is $\phi_1=\phi_2=\phi_3$ and
$\phi_4=\phi_5=\phi_6=\phi_7$. In terms of $\psi$'s these read:
$\psi_1=\psi_2=0$, $\psi_4=\sqrt{3/5}\,\psi_3$,
$\,\psi_5=\sqrt{2/5}\,\psi_3$ and $\psi_6=\sqrt{2/7}\,\psi_3$. There
are only three independent scalar fields: $P$, $\psi_3$ and
$\psi_7$. The system (\ref{eq5})-(\ref{eq6}) has a unique
stationary point $P=\psi_3=\psi_7=0~(=\phi_a)$ around which the
linearized eqs. (\ref{eq6}) are
\beqa
\Box P&\!\!\simeq\!\!& -\frac{2}{5}\Lambda P\label{linearp},\\
\Box \psi_3&\!\!\simeq\!\!&\frac{2}{105\sqrt{2}}\Lambda\big(8\sqrt{2}\,\psi_3
-3\sqrt{21}\,\psi_7\big)\label{linearx3},\\
\Box \psi_7&\!\!\simeq\!\!&\frac{6}{35\sqrt{21}}\Lambda\big(-8\sqrt{2}\,\psi_3
+3\sqrt{21}\,\psi_7\big)\label{linearx7}.
\eeqa

{}From the trace of the Einstein equations it follows that in full
generality $\dot H<0$; thus, if initially $H<0$, the system can only
evolve towards the physical singularity $H\to -\infty$.
Remembering that in the FLRW metric $\Box P=-(\ddot P+3H\dot P)$,
a negative, and
large in modulus, $H$ is equivalent to the presence of an anti-friction
force acting on the fields, that makes unstable any equilibrium point.

The phase space $(P,\dot P)$ is illustrated in
Fig.~\ref{kk1},
in Poincar\'e projection. Along trajectories pointing to North (N),
$P\to +\infty$, $\dot P\to +\infty$ or to South (S),
$P\to -\infty$, $\dot P\to -\infty$, the Hubble parameter is negative, $H<0$.
Since $H$ has two opposite values, at each point of the phase space there
are two trajectories, but, for simplicity we have shown only one of them
in the figure.
Note also that the points N and S
are reached not only when the initial value of $H$ is negative;
since $\dot H<0$ for scalar field dynamics,
also starting with $H>0$ many solutions will turn to $H<0$.
In all these cases
the Universe reaches a physical singularity ($H,\dot H\to -\infty$) in a
finite time interval. The remaining possibility (note that the West point (W)
 lies in the
forbidden region of the phase-space) is given by the trajectories
pointing to East (E), $P\to +\infty$, $\dot P\to 0$. In this case $H\to 0^+$
for
$t\to +\infty$. An example of such a solution is shown in Fig.~\ref{kk2}.

It is of relevant importance that also the attractor trajectory
in Fig.~\ref{kk1} has the same properties. Its analytic expression can
be found by looking for solution of the form $\dot{P}=A{e}^{mP}$
in the limit $P\to+\infty$. Under condition
$\psi_3=\psi_7=\dot{\psi}_3=\dot{\psi}_7=0$ (the condition is
always met asymptotically when $H>0$ since it turns out to be a
stable focus point of the linear system
(\ref{linearx3})-(\ref{linearx7})) and inserting the trial solution
in the field equation (\ref{eq5})-(\ref{eq6}) one finds that the
evolution of the scale factor $a(t)$ for flat FLRW universe is
\begin{eqnarray}\label{eq11abc}
a&=&a_0\left[1-mA{e}^{mP_0}(t-t_0)\right]^p\,,\qquad{\rm and}\nonumber\\
P&=&P_0-\frac{1}{m}\ln\left[1-mA{e}^{mP_0}(t-t_0)\right]\,,\qquad
{\rm with}\\
p&=&\frac{2}{\beta^2}\,,\qquad m=-\frac{\beta}{2}\,,\qquad
A=\left(\frac{2\beta^2\Lambda}{6-\beta^2}\right)^{1/2}\,,\nonumber
\end{eqnarray}
where $t_0$ is an integration constant, and $a_0=a(t_0)$, $P_0=P(t_0)$.
In the present case ($n=8$), along the
attractor, the physical evolution of the system is described by
the leading terms (at $t\to\infty$) in (\ref{eq11abc}),
\beq\label{eq13}
a\approx\left(\frac{t}{t_i}\right)^{5/4}~~\mbox{and}~~P
\approx\sqrt{\frac{5}{2}}\ln t,
\eeq
where $t_i=(-mA{e}^{mP_0})^{-1}$ and $a_0=1$. This is an
inflationary stage, though of a rather mild kind (power law
inflation), see Fig.~\ref{kkattr}.

To summarize the $SU(3)$ case, it is possible to have power law
inflation with the scalars $\psi_3$ and $\psi_7$ dissipating
their kinetic energy on the plateau of their effective potential,
but then the mean volume (proportional to ${e}^{nF}$) of the
internal space increases. In fact, there exists in the phase
space an attractor solution along which the leading-term
evolution is given by (\ref{eq13}). The logarithmic growth of
$P$ (or $F$) seems rather slow but actually it is not slow
enough to have a negligible effect on physics in our external
world. Primordial nucleosynthesis is very stringent in this
respect \cite{infl}: the ratio of the values of the radius
$b(t)$ of the internal space at the nucleosynthesis epoch
($b=b_{\rm N}$ at $t=t_{\rm N}$) and today ($b=b_{0}$ at
$t=t_{\rm now}$) is constrained by $0{.}99\leq b_{\rm
N}/b_0\leq1{.}01$. Since $b=L\exp
F=L\exp\left[(P+q)/\alpha\right]$ this means, for
$P$ given in (\ref{eq13}), that
\beq\label{eq14}
-0{.}01005\leq\frac{2}{\alpha\beta}\ln\frac{t_{\rm N}}{t_{\rm now}}
\leq0{.}00995.
\eeq
One finds $\ln(t_{\rm N}/t_{\rm now})=-34{.}54$, for
$t_{\rm N}\approx 10^2$~sec and $t_{\rm now}\approx 10^{17}$~sec, thus
the solution does not meet the constraint, and the internal
space evolves far too rapidly. Furthermore, the fact that the
power law inflation corresponds to the asymptotic attractor of
the phase space means that the universe never stops inflating
and never enters the hadron dominated era. Thus the geometrical
inflation from the extra dimensions encounters the graceful exit
problem and the latter should be solved by an independent
physical mechanism (e.g. inserting a counterterm in the dilaton potential).

Although the nucleosynthesis constraint excludes the solution
(\ref{eq13}), it is worth noticing that the growth of the radius
of the internal space needs not make it observable at the end of
inflationary epoch. Assuming traditionally \cite{infl1} that
satisfactory inflationary evolution corresponds to $N=60$
$e$-foldings, $N=\ln (a_{\rm final}/a_{\rm
initial})$, one finds from (\ref{eq13}) $N=p\ln
\left(t_{\rm final}/t_{\rm
initial}\right)$ and the growth of the extra dimensions is
\beq\label{eq15}
\frac{b_{\rm final}}{b_{\rm
initial}}=e^{\beta N/\alpha}={e}^{N/5}\approx 1{.}6\cdot 10^5.
\eeq

Unless $b_{\rm initial}$ is many orders of magnitude larger than
the Planck length, it remains unobservable after 60 $e$-foldings
of the inflationary evolution. The duration of the inflationary
era is $t_{\rm final}/t_{\rm initial}={e}^{48}\approx
10^{21}$, an acceptable value provided inflation starts soon after the
Planck era.

As the second case one takes the internal space as being either
isometric to a ``squashed" 3-sphere or the product of two
spheres, ${\cal B}=S^3\times S^3$.

The common feature of all the models we have considered is that
all the subspaces whose product composes the internal space
isotropize separately. For instance, if $\cal B$ is the product of
$N$ three-spheres, $n=3N$, then each sphere isotropizes while
the scale factors for different spheres will diverge in general.
The divergence is due to the constraint $\sum_a\phi_a=0$; it is
easily seen in the case of $N-1$ spheres having equal scale
factors, $\phi_a=\phi$ for $a=1,\ldots,n-3$ and
$\phi_a\equiv-(N-1)\phi$ for $a=n-2,n-1,n$, then the last sphere
evolves in the apposite direction to the others. Thus the
internal space remains anisotropic although the degree of
anisotropy decreases.

The effective potential $V(P,\phi)$ is now unbounded from below
and has a global maximum at $P=\phi_a=0$, the extremum point
belongs to the constraint hypersurface. The qualitative features
of the evolution can be summarized as follows: a) for a single
sphere, the axially symmetric (oblate) and isotropic
configurations are energetically favourable; however the process
of symmetrization and the dilaton expansion or contraction occurs
with similar time scales; b) if the initial conditions for $P$
are such that  first an inflationary phase occurs, i.e. if $P$
starts near $P=0$ with negligible velocity, then the duration of
the phase is not increased by anisotropic initial conditions; c)
for the product of two or more spheres each of them evolves
towards the ``round" sphere with different radius (see above);
d)~in addition to the singular solutions $H\to-\infty$, there
are always generic solutions exhibiting the power law inflation
in four dimensions, $H=p/t$, $p>1$, corresponding to an
attractor trajectory in the phase space $(P,\dot{P})$.

Since any initial anisotropy of each 3-sphere tends to vanish
(Fig.~\ref{kks3_2}),
let us consider the case of a single isotropic sphere. Using the
same method as in the $SU(3)$ case one finds an attractor
solution in the limit $P\to+\infty$ (Fig.~\ref{kks3_1}). The leading terms
in the solution (\ref{eq11abc}) are, now,
\beq\label{eq16}
a(t)\sim t^{5/3}~~\mbox{and}~~P\approx\sqrt{\frac{10}{3}}\ln t.
\eeq
The external world expands even faster than in the case of
$SU(3)$ space, therefore the sufficient inflation, $N=60$
$e$-foldings, is achieved in a shorter time, $t_{\rm final}/t_{\rm
initial}\approx {e}^{N/p}\approx {e}^{36}\approx 10^{15}$.
Also the internal space expands faster and in this inflationary
period its growth is
\beq\label{eq17}
\frac{b_{\rm final}}{b_{\rm initial}}\approx
\exp\left({\frac{2N}{3p}}\right)\approx {e}^{24}\approx 10^{10},
\eeq
an acceptable factor. Once again the nuclesynthesis constraint
excludes this attractor solution as incompatible with
observations: the constraint (\ref{eq14}) is replaced by
\beq\label{eq18}
-0{.}01005\leq \frac{2}{3}\ln \frac{t_{\rm N}}{t_{\rm now}} \leq
0{.}00995
\eeq
and the latter cannot be satisfied, unless an exit mechanism is provided.

Finally, we compare our results with a recent work by Levin
\cite{jl}. The difference is that Levin does not use the
conformal factor ${e}^{-nF}$ to construct the physical metric
field in four dimensions and the internal space in that work is
a multidimensional torus. The comparison clearly shows the
impact of the factor and of the curvature of $\cal B$ on the
evolution of the external world. While we assume that $({\cal
M},{\bf g})$ is
the physical spacetime, Levin takes $({\cal M},\hat{\bf g})$ with
$\hat{g}_{\alpha\beta}= {e}^{-nF}g_{\alpha\beta}$. The
radius of the internal space is defined in the same way here and
in \cite{jl}, $b=\hat{b}={e}^F$ ($L=1$ for simplicity). If
$g_{\alpha\beta}$ is the metric of flat FLRW spacetime than the
cosmic scale factors are related by
\beq\label{eq19}
\hat{a}=\exp\left(-\frac{n}{2}F\right)a=b^{-\frac{n}{2}}a.
\eeq
In the case of flat $n$-torus there is no effective potential
and the evolution of $\hat{a}$ is driven solely by the time
derivatives of the scalar field $b$ (``kinetic inflation").
Assuming $\Lambda=0$ it was found that \cite{jl}
\begin{eqnarray}\label{eq20}
\hat a&=&c_1\,(1-\tau/\tau_0)^{-k}\,,\qquad
k\equiv\frac{-3+\sqrt{3n^2+6n}}{3(n+3)}\qquad\mbox{and}\nonumber\\
\hat b&=& b=c_2\,(1-\tau/\tau_0)^w\,,\qquad
w\equiv\frac{n+\sqrt{3n^2+6n}}{n(n+3)}\,,
\end{eqnarray}
where $c_1$, $c_2$ and $\tau_0$ are constants and $\tau$ is the
proper time for the metric $\hat{g}_{\alpha\beta}$ related by a
conformal transformation to the proper time $t$ for $g_{\alpha\beta}$,
\beq\label{eq21}
{\rm d}\tau=\exp\left(-\frac{n}{2}F\right){\rm
d}t=b^{-n/2}{\rm d}t\,.
\eeq
This is an accelerating (``inflationary") solution, starting
from a regular state at $\tau=0$ and evolving towards a future
singularity in $b(\tau)$ at $\tau_0$. In terms of $({\cal
M},{\bf g})$ viewed
as the physical spacetime, this solution has a quite different
interpretation. It can be obtained as a solution to (\ref{eq5})
and (\ref{eq6}) for $\Lambda=V(P,\psi)=\psi_i=0$, $i=1,\ldots,n$.
Then it reads
\beq\label{eq22}
a= c_3\,(\tilde{t}-t)^{1/3}~~\mbox{and}~~P=\sqrt{\frac{2}{3}}
\ln(\tilde{t}-t)
\eeq
with constant $c_3$ and $\tilde{t}$. One can easily check that
(\ref{eq20}) and (\ref{eq22}) are related via transformations
(\ref{eq19}) and (\ref{eq21}). Now expansion has turned into
contraction and a future singularity appears also in four
dimensions. There exists also a slowly growing solution
\cite{jl}, $\hat{a}\propto \tau^r$ with $0<r<1$; in the metric
$g_{\alpha\beta}$ it corresponds to $a\propto t^q$ with $0<q<1$.
Since in our opinion the physical metric is $g_{\alpha\beta}$ and not
the conformally related $\hat{g}_{\alpha\beta}$
\cite{DD,Maeda,SC,Cho19871992}, we conclude that in the model with flat
internal space and vanishing cosmological constant,
an inflationary evolution of the external world is not present.

It is worth noticing, however, that the contracting solution (\ref{eq22})
is accelerating ($\dot{a}<0$ and $\ddot{a}<0$), i.e., it
represents {\em deflation}. Recently it was shown \cite{gave}
that deflation may also solve the horizon, flatness and
amplification of perturbation problems
of the big-bang cosmology. In this sense, also the toroidal internal
space  can be considered among the viable models of the primordial
universe.

In conclusion, anisotropic or isotropic cosmological evolution
driven by the curvature in the higher internal dimensions turns
out to be generically of the form of power law inflation. The
essential ingredients are a curved internal space (this was
shown for the case of $SU(3)$ group manifold and 3-spheres) and
a positive cosmological constant. There is no matter in the
multidimensional world and gravity is the only interaction.
Matter appears only after dimensional reduction in the form of
self-interacting scalar fields of geometrical origin. This approach
allows one to avoid introducing {\em ad hoc\/} matter and an
inflaton field. The dynamics of the four-dimensional world is
entirely determined by geometry of the internal space. The
inflationary epoch (attractor solution) is long enough to solve
the horizon and flatness problems of the standard cosmology. The
power-law inflation must be terminated by an unknown mechanism
(the graceful exit problem). Furthermore, this or another
mechanism should stabilize the internal space at sufficiently
small size (the growth of the space should be stopped
approximately simultaneously with the end of the inflationary
epoch) and turn the scalar fields into ordinary elementary
particles. These mechanisms are beyond our model. Nevertheless
it is interesting to find purely geometrical inflation without
matter, fine-tuning and special potentials.

Details of all the calculations will be published elsewhere
\cite{glsad}.

\vspace{.2cm}
\leftline{\large\bf Acknowledgement}
\vspace{.2cm}

We are grateful to Prof. Andrzej Staruszkiewicz for helpful
comments and discussions. The work of Z.A.~G., L.M.~S. and A.~D.
was partially supported by grants of the Polish Committee for
Scientific Research.
\eject

\centerline{\bf Figures}
\begin{enumerate}
\item\label{kk1}
The phase-space (in Poincar\'e projection) of the dilaton $P$,
in the $SU(3)$ maximally symmetric case.  $P$ is proportional to the logarithm
of the mean radius of the internal space.
The most interesting
feature is the {\em attractor trajectory} pointing to E, with $P\to
+\infty$, $\dot P\to 0$.

\item\label{kk2}
The whole dynamics of the universe along the attractor trajectory shown
in Fig.~{\ref{kk1}}.
{\bf(\ref{kk2}a)} The Hubble parameter $H$ goes asymptotically to zero.
{\bf(\ref{kk2}b)} The expected behaviour along the attractor $P\to +\infty$,
$\dot P\to 0$.
{\bf(\ref{kk2}c)} All the other fields of the model go to the stationary
point $\phi_1=\phi_2=\phi_3=0$,
$\phi_4=\phi_5=\phi_6=\phi_7= 0$.

\item\label{kkattr}
Power law inflation in the external spacetime  of scale factor $a(t)$, is
naturally present
in the $SU(3)$ internal space case. In fact, for a large set of
initial
conditions, once the attractor trajectory has been reached, it happens that:
{\bf(\ref{kkattr}a)} $H\to 0$,
{\bf(\ref{kkattr}b)} $a(t)\propto t^{5/4}$. Of course, different solutions do
not reach the attractor at the same time.

\item\label{kks3_2}
In the $S^3$ internal space, complete isotropy is quickly
achieved.  {\bf(\ref{kks3_2}a)} The initial degree of anisotropy,
measured by $\psi_2\neq 0$, in a comparison to $P$, is quickly
swept away.  {\bf(\ref{kks3_2}b)} The corresponding runs of $H$
overlap for most of the inflationary stage, i.e. for large
$a$ and $P$, when $H$ is almost constant. Note also the plateau
at $a\geq 1$; a larger initial anisotropy does not improve the
expansion that results to be of only a couple of $e$-foldings.

\item\label{kks3_1}
The phase-space (in Poincar\'e projection) of the dilaton $P$,
in the $S^3$ case.  In this case also there is an attractor
trajectory $P\to +\infty$, $\dot P\to 0$.  Along it, $a\propto
t^p$, with $p\approx 1{.}67$, and again we are in presence of a
power law inflation.

\end{enumerate}
\end{document}